\documentclass[aps,prl,twocolumn,showpacs,groupedaddress]{revtex4}  
\usepackage{amsmath}
\usepackage{epsfig}
\usepackage{setspace}
\usepackage{textcomp}

\usepackage{graphicx}  
\usepackage{dcolumn}   
\usepackage{bm}        
\usepackage{amssymb}   
\usepackage{changebar}  
\begin{document}



\title{Phase Space Approach to Solving the Time-independent Schr\"odinger Equation}
\author{Asaf Shimshovitz and David J.~Tannor}                             
\affiliation{Department of Chemical Physics, Weizmann Institute of Science, Rehovot, 76100 Israel}
\date{\today}

\begin{abstract}
We propose a method for solving the time independent Schr\"odinger equation based on the von Neumann (vN) lattice of phase space Gaussians. By incorporating periodic boundary conditions into the vN lattice [F. Dimler et al., New J. Phys. 11, 105052 (2009)] we solve a longstanding problem of convergence of the vN method.  This opens the door to tailoring quantum calculations to the underlying classical phase space structure while retaining the accuracy of the Fourier grid basis.  The method has the potential to provide enormous numerical savings as the dimensionality increases. In the classical limit the method reaches the remarkable efficiency of 1 basis function per 1 eigenstate. We illustrate the method for a challenging two-dimensional potential where the FGH method breaks down.  
\end{abstract}

\maketitle

The formal framework for quantum mechanics is an infinite dimensional
Hilbert space. In any numerical calculation, however, a wave function
is represented in a finite dimensional basis set and therefore the 
choice of basis set determines the accuracy. The optimal basis set should
combine accuracy and flexibility, allowing a small number of basis
functions to represent the wave functions even in the presence of
complex boundary conditions and geometry. Unfortunately, these two
criteria ---accuracy and efficiency--- are usually in conflict, and globally accurate methods
\cite{fourier_method,fgh,wei}
lack the flexibility of local methods
\cite{dgb,garashchuk,davis_and_heller,Hamilton}.     
For example, in the 
pseudospectral Fourier grid method the wave function is represented by its values on a finite number of evenly spaced grid points. Due to the Nyquist sampling theorem, this allows for an exact representation of the wave provided the wavefunction is band limited with finite support\cite{whittaker,nyquist,shannon}. However, the non-local form of the basis functions in momentum space leads to limited efficiency. 
On the other hand, in the von Neumann basis set \cite{brixner,von_neumann} each basis function is
localized on a unit cell of size $h$ in phase space. However, despite the formal completeness
of the vN basis set\cite{perelomov}, attempts to utilize
this basis in quantum numerical calculations 
have been plagued with numerical errors\cite{davis_and_heller,poirier}.

In this paper we establish a precise mathematical formalism for the vN basis on a truncated phase space. By using periodic boundary conditions in the vN basis, as introduced in the seminal work by Dimler et al. \cite{dimler}, the method achieves Fourier accuracy with Gaussian flexibility.  This allows one to tailor the basis in quantum eigenvalue calculations to the underlying classical phase space structure, with the potential for enormous numerical savings. The efficiency of the method relative to the Fourier grid rises steeply with dimensionality, defeating exponential scaling. In the classical limit the method reaches the remarkable efficiency of 1 basis function per 1 eigenstate.



The von Neumann basis set \cite{von_neumann} is a subset of
the ``coherent states'' of the form:
\begin{eqnarray}
g_{nl}(x)=\left(\frac{2\alpha}{\pi}\right)^\frac{1}{4}\exp\left(-\alpha(x-na)^{2}-il\dfrac{2\pi\hbar}{a}(x-na)\right)
\end{eqnarray}%
where $n$ and $l$ are integers. Each basis function is a Gaussian centered
at $(na,\frac{2\pi l}{a})$ in phase space. 
The parameter
$\alpha=\frac{\sigma_p}{2\sigma_x}$ controls the FWHM of each Gaussian in $x$ and
$p$ space. Taking $\Delta x =a, \Delta p
=h/a$ as
the spacing between neighboring Gaussians in $x$ and $p$ space
respectively, we note that $\Delta x \Delta p=h$
so we have exactly one basis function per unit cell in
phase space. As shown in \cite{perelomov} this  implies completeness in the Hilbert space.

The ``complete'' vN basis, where $n$ and $l$ run over all integers, spans the
infinite Hilbert space. In any numerical calculation, however, $n$ and
$l$ take on a finite number of values, 
producing $N$ Gaussian basis functions $\lbrace g_i(x)\rbrace$, $i=1...N$. 
Since the size of one vN unit cell is
$h$, the area of the truncated vN lattice is given by   
$S^{\rm vN}=Nh$.  

\begin{figure}[h]
\begin{center}
\includegraphics [width=4cm]{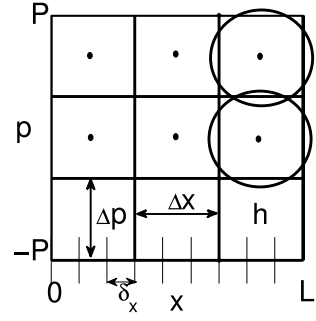}
\end{center}
\vspace{-.6cm}
\begin{center}
\caption{\footnotesize{$N=9$ coordinate grid points and $N=9$ vN unit cells span
the same area in phase space,$S=Nh$. The vN basis
functions are Gaussians located at the center of each unit
cell. }}
\label{VN}
\end{center}
\vspace{-1.2cm}
\end{figure}

The pseudospectral Fourier method (also known as the sinc Discrete Variable Representation \cite{Miller}) is a widely used tool in quantum simulations \cite{day1,goldfield,hayes,josh}. In this method a function $\psi(x)$
that is
periodic in $L$ and band limited in $K=\frac{P}{\hbar}$ can be written
in
the following form: 
$\psi(x)=\sum_{n=1}^{N} \psi(x_n)\theta_{n}(x)$,
where $x_n=\delta_x(n-1)$, and
$\delta_x=\frac{\pi\hslash}{P}=\frac{L}{N}$. 
The basis functions $\lbrace\theta_n(x) \rbrace$ are given by 
\cite{tannor_book}:
\begin{eqnarray}
\theta_n(x)=\sum_{j = \frac{-N}{2} + 1}^{\frac{N}{2}}\frac{1}{\sqrt{LN}}\exp\left(\frac{i2\pi j}{L}(x-x_n)\right),
\end{eqnarray}
which can be shown to be sinc functions that are periodic on the domain $[0,L]$ \cite{supp}.
The set $\lbrace\theta_i(x) \rbrace$ $i=1,..,N$ spans a
rectangular shape in phase space with area of $S^{\rm
FGH}=2LP=2L\frac{\pi\hbar}{\delta_x}=Nh$.
Thus $N$ unit cells in the vN lattice and $N$ grid points in the Fourier method cover the same rectangle with an area in phase space of: 
\begin{eqnarray}
S^{\rm vN}=S^{\rm FGH}=Nh
\label{S1}
\end{eqnarray}
(Fig. \ref{VN}). This suggests that $N$ vN basis functions confined to
this area will be equivalent to the Fourier basis set.  
Unfortunately, the attempt to use $N$ Gaussians as a basis set for the
area in eq.(\ref{S1}) (Fig. \ref{VN}) is unsuccessful, a consequence of the
Gaussians on the edges protruding from the truncated space.
However, by combining the Gaussian and the Fourier basis functions we
can generate a ``Gaussian-like'' basis set that is confined to the
truncated space. We use the basis sets $\lbrace g_i(x) \rbrace$ and
$\lbrace\theta_i(x) \rbrace$ to construct a new basis set, $\lbrace
\Tilde g_i(x) \rbrace$: 
\begin{eqnarray}
\Tilde g_m(x)=\sum_{n =  1}^{N} \theta_n(x)g_m(x_n)
\label{dvn}
\end{eqnarray}
for $m=1,...,N$. The new basis set is in some sense, the Gaussian functions with periodic boundary conditions. We can write eq.(\ref{dvn}) in matrix notation as: $\Tilde G=\Theta G$
where $G_{ij}=g_j(x_i)$ 
By taking the width parameter $\alpha=\frac{\Delta p}{2\Delta x}$ we
can guarantee that the pvN functions have no linear dependence and
that the matrix $G$ is invertible, that is $\Tilde G G^{-1}=\Theta$.
The invertibility of $G$ implies that both bases span the same space. 

The representation of $\vert \psi \rangle$ in the pvN basis set is given by:
\begin{eqnarray}
\vert \psi \rangle=\sum_{m =  1}^{N} \vert\Tilde g_m\rangle a_m.
\label{psi}
\end{eqnarray}
To find the coefficients $a_m$ we first define the overlap matrix, $S$: 
\begin{eqnarray}
S_{ij}&=& \langle\Tilde g_i \vert \Tilde
g_j\rangle=\int_{0}^{L} \Tilde g_i^{*}(x)\Tilde g_j(x)dx
\nonumber \\
&=&\sum_{n =  1}^{N}\sum_{m =  1}^{N} g_i^{*}(x_n)
g_j(x_m)\int_{0}^{L} \theta_n^{*}(x)\theta_m(x)dx
\nonumber \\
&=&\sum_{n =  1}^{N} g_i^{*}(x_n)g_j(x_n)
\label{ip}
\end{eqnarray}
or
\begin{equation}
S=G^{\dag}G.
\label{S}
\end{equation}
Using the completeness relationship for non-orthogonal bases, 
$\vert \psi \rangle$ can be expressed as
\begin{equation}
\vert \psi \rangle=\sum_{n =  1}^{N} \sum_{m =  1}^{N}
\vert\Tilde g_m \rangle (S^{-1})_{mn} \langle\Tilde
g_n\vert\psi\rangle.
\label{psi2}
\end{equation}
Comparing with eq.(\ref{psi}) we find that $a_m=\sum_{n =  1}^{N} (S^{-1})_{mn} \langle\Tilde g_n\vert\psi \rangle$
and $\langle\Tilde g_i\vert\psi \rangle=\sum_{n =  1}^{N} g_i^{*}(x_n)\psi(x_n)$.

Although the periodic von Neumann (pvN) and the Fourier methods span the same rectangle in phase
space, the localized nature of the basis functions in the pvN method can lead to significant advantages. In particular, if $\vert\psi\rangle$ has an irregular phase space shape we may expect that some of the pvN basis functions will fulfill the relation: $\langle\Tilde g_j|\psi\rangle=0$, $j=1,...,M$. Due to the non-orthogality of the basis we cannot simply eliminate the states $\tilde{g}_j$, since the \textit{coefficients} of $\tilde{g}_j$ may include contributions from remote basis functions, but 
we can take advantage of the vanishing overlaps by defining a bi-orthogonal von Neumann basis (bvN) $\{{b}_i(x)\}$. 
\begin{eqnarray}
\vert b_i\rangle=\sum_{j =  1}^{N} \vert\Tilde g_j\rangle (S^{-1})_{ji}
\label{fvn}
\end{eqnarray}
or in matrix notation: $B=\Tilde GS^{-1}.$ 
Inserting eq.\ref{fvn} into eq.\ref{psi2}, $\vert \psi \rangle$ can be
written as 
\begin{eqnarray}
\vert \psi \rangle=\sum_{n =  1}^{N} \vert b_n\rangle c_n=\sum_{n =  1}^{N} \vert b_n\rangle \langle \Tilde g_n\vert\psi\rangle. 
\end{eqnarray}
By assumption, $M$ of the coefficients are zero, hence in order to
represent $|\psi\rangle$ in the bvN basis set  we need only $N'=N-M$
basis functions.
Note that the bvN and pvN are bi-orthogonal bases, meaning that each
set taken by itself is non-orthogonal but they are orthogonal to each other. This
can be shown easily by:
\begin{eqnarray}
\langle \Tilde g_i\vert b_j\rangle&=&\sum_{n =  1}^{N} g^{*}_i(x_n)f_j(x_n)
\nonumber\\
&=&\sum_{m =  1}^{N} \sum_{n =  1}^{N} g^{*}_i(x_n)g_m(x_n)(S^{-1})_{mj}
\nonumber\\
&=&\sum_{m =  1}^{N} S_{im}(S^{-1})_{mj}=\delta_{ij}.
\end{eqnarray}
For many practical applications the full knowledge of
the basis wavefunctions is unnecessary: we need only the value of
the basis functions at the sampling points. For example the evaluation
of Hamiltonian matrix elements can be performed explicitly by:

\begin{eqnarray}
H^{\rm pvN}_{ij}&=&\langle\Tilde g_i\vert H\vert \Tilde g_j\rangle
\nonumber\\
&=&\sum_{m =  1}^{N}\sum_{n =  1}^{N} g_i^{*}(x_m)\langle\theta_m\vert H\vert\theta_n\rangle g_j(x_n)
\nonumber\\
&=& \sum_{m =  1}^{N}\sum_{n =  1}^{N} g_i^{*}(x_m)H^{\rm FGH}_{mn}g_j(x_n) 
\end{eqnarray}
and similarly:
\begin{eqnarray}
H^{\rm bvN}_{ij}=\sum_{m =  1}^{N}\sum_{n =  1}^{N}
b_i^{*}(x_m)H^{\rm FGH}_{mn}b_j(x_n) 
\end{eqnarray}
where $H^{\rm FGH}=V^{\rm FGH}+T^{\rm FGH}$ and the potential and the
kinetic matrix are given by: $V^{\rm FGH}_{ij} \approx V(x_i)\delta _{ij}$
and
\begin{eqnarray}
T^{\rm FGH}_{ij} =\frac{\hslash^{2}}{2M}
\begin{cases}
  \frac{K^{2}}{3}(1+\frac{2}{N^{2}}),  & \mbox{if} \quad i = j \\
  \frac{2K^{2}}{N^{2}}\frac{(-1)^{j-i}}{\rm {sin}^{2}(\pi\frac{j-i}{N})}, & \mbox{if} \quad i \neq j
\end{cases}
\end{eqnarray}
\cite{tannor_book2}. The eigenvalue problem in a non-orthogonal basis set becomes 
$HU=sUE$; in the pvN basis set  $s$ is given by 
eq. (\ref{S}) and in the bvN basis set $s$ is given by: 
\begin{equation}
B^{\dag}B=S^{-1}G^{\dag}GS^{-1}=S^{-1}.
\end{equation} 
Diagonalization should give accurate results for all wavefunctions
localized to the classically allowed region of the rectangle. 
Note that in the multidimensional implementation, the $S^{-1}$ matrix required in Eq.(9) can be constructed separately for each dimension. As a result, the computational effort to construct the bvN basis set is negligible compared with diagonalizing the Hamiltonian. 

\begin{figure}[h]
\begin{center}
\includegraphics [width=7cm]{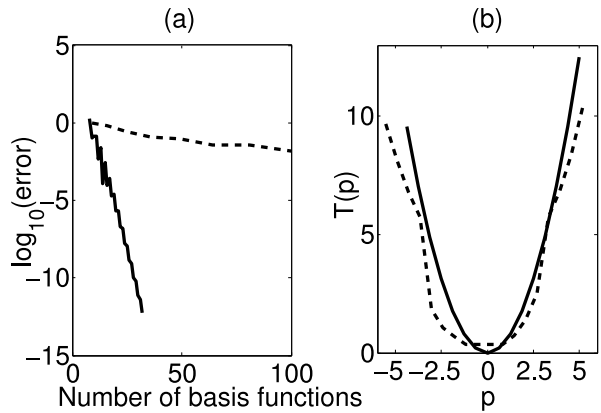}
\end{center}
\vspace{.2cm}
\begin{center}
\caption{\footnotesize{ (a) Error in the $7^{\rm{th}}$ eigenvalue of the harmonic oscillator as a function of basis set size for vN(dashed) and pvN(solid). (b) Kinetic energy spectra using 16 basis functions. vN(dashed), pvN(solid).}}
\label{harmonic}
\end{center}
\vspace{-.8cm}
\end{figure}

As a numerical test of the pvN basis set we studied the standard example of the
harmonic oscillator $V(x)=\frac{m\omega^{2}x^{2}}{2}$ in units such
that $m=\hbar=\omega=1$. We calculated the seventh excited energy using different number of pvN and conventional Gaussian basis functions. 
In the Gaussian basis set the Hamiltonian and the overlap  matrices
were calculated analytically as: $H_{ij}=\langle
g_i|H|g_j\rangle=\int_{-\infty}^{\infty}g^{*}_i(x)[-\frac{d^{2}}{dx^{
2 }}+V(x)]
g_j(x)dx$
and
$S_{ij}=\langle
g_i|g_j\rangle=\int_{-\infty}^{\infty}g^{*}_i(x)g_j(x)dx$. 
The results, shown in Fig. \ref{harmonic}(a), show the superiority of the pvN basis set compared to
the standard Gaussian basis set. In fact, the results obtained with
the pvN basis set are exactly as accurate as in the Fourier grid
method. The kinetic energy spectra in Fig. \ref{harmonic}(b) show that the pvN has a perfect quadratic dependence while the vN spectrum is highly flawed. 

\begin{figure}[h]
\begin{center}
\includegraphics [width=7.5cm]{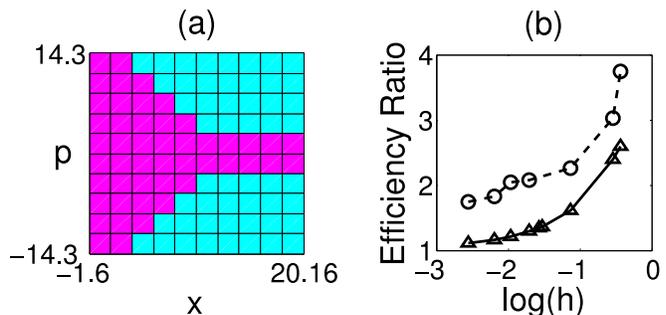}
\end{center}

\vspace{.3cm}
\begin{center}
\caption{\footnotesize{(a). Phase space area spanned in
the bvN method (magenta) and in the pvN (or FGH) method (full rectangle) for
Morse. (b) Efficiency ratio (defined as number of basis functions per converged eigenstates) of the pvN (solid) and FGH (dashed) methods for the Morse potential as
function of $\hbar$.}}
\label{phase}
\end{center}
\vspace{-.3cm}
\end{figure}

In the bvN basis set we are able to remove some of the basis functions
and construct lower dimensional $H^{\rm bvN}$ and $S^{\rm bvN}$
matrices without losing accuracy.  In order to test this claim, we
calculated numerically the eigenenergies of the Morse oscillator
$V(x)=D(1-e^{-\beta x})^{2}$ 
by using both the FGH and bvN basis sets.
The
Morse parameters were taken to be $D=12$, $m=6$, $\beta=0.5$ and
$\hbar=1$. For FGH, 100  grid points between $[-1.6,20.1]$ were
required to get 4 digits of accuracy in energy for all 24
bound states. By using the bvN basis functions (constructed from
10$\times$10 vN functions with $\alpha=0.5$)
we obtain the same 4 digit accuracy with only 48 basis functions.
This is demonstrated graphically in Fig. \ref{phase} (a). The
figure shows the phase space representation of 100 evenly grid points.
Although it requires 100 pvN basis functions to span this area in phase space,
due to the flexibility of the bvN basis set we can 
suffice with just the basis functions in the  classically  allowed
region (magenta squares).
 
\vspace{.5cm}


%

\begin{figure}[h]
\begin{center}
\includegraphics [width=7cm]{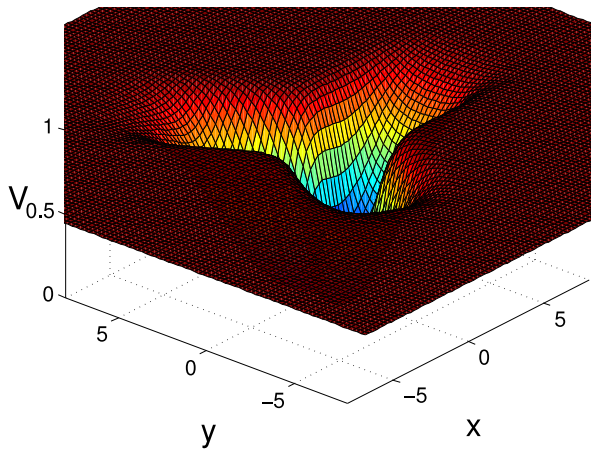}
\end{center}
\begin{center}
\caption{\footnotesize{The triangle potential: a two dimensional test case for the bvN method.}}
\label{coulomb}
\end{center}
\vspace{-1cm}
\end{figure}
\begin{figure}[h]
\begin{center}
\includegraphics [width=8cm]{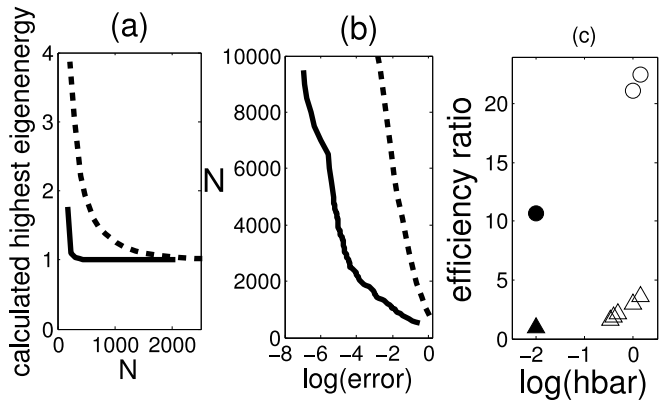}
\end{center}
\vspace{.2cm}
\begin{center}
\caption{\footnotesize{Triangle potential results for bvN(solid) and FGH(dashed) (a) The calculated highest eigenenergy as a function of basis set size $N$. (b) The accuracy of the calculated highest eigenenergy as a function of basis set size $N$. (c) Efficiency ratio of the pvN ($\bigtriangleup$) and FGH ({\large$\circ$}) methods as a function of $\hbar$. The $\blacktriangle$ (pvN) and {\large$\bullet$} (FGH) signify that the value is an approximation to the $\hbar\rightarrow 0$ value, given by the ratio between the size of the phase space spanned by the basis and the classical phase space.
}}
\label{ratio}
\end{center}
\vspace{-1cm}
\end{figure}

The ability to localize a bvN function at a specific point in phase
space results in the remarkable concept of 1 basis function per 1
eigenstate. This means that in order to calculate $N$ eigenenergies we
need only $N$ basis functions. Obviously, such one per one efficiency,
if reachable, will be the ideal efficiency for any basis set. 
In order to test the ability of the bvN method to reach the
ideal efficiency we examined the Morse potential and
looked for the smallest basis that provides 12 digits of accuracy for all the eigenvalues up to $E=11.25$. The bvN method indeed tends to the ideal
efficiency in the classical limit $\hbar\rightarrow 0$
(Fig. \ref{phase}b). This remarkable result is unique for methods based on phase space localization \cite{poirier2}.
The true power of the method is in the application to higher dimensional systems. 
As an illustration, consider the potential
$V(r,\theta) = (1-\exp(-\alpha(\theta) r^2))^2$  
where
$\alpha = ((1-\cos(3\theta))/4)^2+0.05$.   
This 3-fold symmetric potential (Fig. (\ref{coulomb})), which is a realistic model for a system of three identical particles and fixed hyperradius, is quite challenging for the FGH method. Taking $m=96$, $\hbar=1$ gives 760 states below $E=0.996$.  In order to get two digits of accuracy for all those states one needs $\sim11000$ FGH grid points, while with the bvN basis set convergence is achieved with only 1500 basis functions. For higher accuracy (four digits), the FGH breaks down completely 
while the bvN method requires fewer than 3000 basis functions (Fig.(\ref{ratio}a-b)). Figure (\ref{ratio}c)) shows again that as $\hbar \rightarrow 0$ the efficiency tends to 1 basis function per 1 eigenstate (because of the size of the calculations we consider just 3 digits of accuracy). In contrast, the FGH efficiency as $\hbar \rightarrow 0$ is determined by the ratio between the classical phase space and the box that contains it, which we calculate to be $\sim10$ for this system using Monte Carlo integration. 
Note that in going from 1-d to 2-d the savings provided by the bvN relative to the FGH method has gone from 2 to 7-10 for qualitatively similarly potentials. This suggests that the relative efficiency of the bvN method increases rapidly with dimension.

To explore the scaling with dimensionality more fully, consider a harmonic oscillator with 1-d classical phase space volume $v$ up to energy $E$. For the $D$-dimensional oscillator, the total phase space volume up to energy $E$ is $V=v^D/D!$ In the classical limit, the total number of states is determined by $V/h^D$ and therefore in this limit the efficiency of pvN relative to FGH is determined by the ratio of the phase space volumes spanned. Defining $a$ to be the area of the box surrounding the 1-d oscillator phase space, the volume of the box surrounding the $D$-dimensional phase space is $A=a^D$ and the ratio of phase space volumes is $S=V/A=s^D/D!$ where $s=v/a=\pi/4$ for the harmonic oscillator. For the Morse, Coulomb and other chemically relevant potentials, the 1-d ratio $s < \pi/4$ and the $D$-dimensional phase space volume scales more slowly than $v^D/D!$ \cite{supp}; these effects combine so that the relative efficiency of the pvN method rises steeply with dimension.  As a result of the  $D!$ in the expression for $V$, the method remarkably defeats exponential scaling. A more detailed analysis \cite{supp} shows that for $D \gg v/h =g$ the method scales polynomially: $V=D^g/g!$ 

Work in progress includes application to vibrational eigenvalue calculations for realistic polyatomic molecules, electronic eigenvalues for multielectron atoms and extension of the approach to the time-dependent Schr\"odinger equation.

This work was supported by the Israel Science Foundation and made possible in part by the historic generosity of the Harold Perlman family. We thank Bill Poirier for helpful discussions.
\vspace{-.6cm}





\end{document}


\title{SUPPLEMENTARY MATERIAL}
\maketitle
\section{THE Dirichlet or periodic sinc function}
In Eqn. (2) of the paper we introduced the pseudospectral functions underlying the FGH/sinc DVR method as a sum of exponentials:
\begin{eqnarray}
\theta_n(x)=\sum_{j = \frac{-N}{2} + 1}^{\frac{N}{2}}\frac{1}{\sqrt{LN}}\exp\left(\frac{i2\pi j}{L}(x-x_n)\right).
\end{eqnarray}
In this section we will show that these pseudospectral functions $\theta(x)$ are periodic sinc functions, proportional to the Dirichlet functions. We will first replace the index $j$ with $k=j+N/2-1$ and define $A\equiv\frac{2\pi(x-x_n)}{L}$ (for simplicity we suppress the subscript $n$ on $A$) to obtain:
\begin{eqnarray}
\theta_n(x)&=&\sum_{k = 0}^{N-1}\frac{1}{\sqrt{LN}}\exp\left(iA(k-N/2+1)\right)
\nonumber \\
&=&\frac{\exp\left(iA(-\frac{N}{2}+1)\right)}{\sqrt{LN}}\sum_{k = 0}^{N-1}\left(\exp\left(iA\right)\right)^k.
\end{eqnarray} 
For $A=2\pi m$, where $m$ is an integer, $\theta_n(x)=\sqrt{\frac{L}{N}}$; for $A\neq 2\pi m$ we use the 
expression for a geometric series $\sum_{k=0}^{N-1}r^k=\frac{1-r^N}{1-r}$ to get:
\begin{eqnarray}
\theta_n(x)&=&\frac{\exp\left(iA(-\frac{N}{2}+1)\right)}{\sqrt{LN}}\frac{1-\exp(iAN)}{1-\exp(iA)}
\nonumber \\
&=&\frac{\exp(\frac{iA}{2})}{\sqrt{LN}}\frac{\sin(\frac{NA}{2})}{\sin(\frac{A}{2})} \nonumber \\
&=& \frac{\exp(\frac{iA}{2})}{\sqrt{LN}}D_N(A)\nonumber, 
\end{eqnarray}
where $D_N(A)$ is the Dirichlet function.

\section{Scaling of the pvN method with dimensionality}
In this section we discuss the scaling of the pvN method with the dimensionality of the system. We begin in Section \ref{sec:ho_volume} by discussing the scaling of the classical phase space volume $V$ up to energy $E$ of a set of $D$ isotropic harmonic oscillators. We show that $V$ increases more slowly than exponential with $D$. 

Since the number of states $G$ is given semiclassically by $G=\frac{V}{h^D}$, and since in the classical limit the pvN method requires only 1 basis function per eigenstate, we find that the scaling of the pvN method is slower than exponential with $D$. 

In Section \ref{sec:efficiency}, we note that semiclassically the efficiency of the pvN relative to the FGH method is (the inverse of) the ratio of the classical phase space volume to that of the enclosing box. For harmonic potentials, this relative efficiency increases faster than exponentially with dimension $D$. For anharmonic potentials, two factors contribute to make the relative efficiency even higher than in the harmonic case.

In Section \ref{sec:number_of_states} we go beyond the semiclassical formulae.  Starting with a fully quantum formula for the number of states of a $D$-dimensional isotropic harmonic oscillator, we show that the scaling with $D$ is sub-exponential, leading in the limit of large $D$ to polynomial scaling, $G=D^g/g!$.

\subsection{Phase space volume $V$ of the $D$-dimensional oscillator}
\label{sec:ho_volume}
The phase space of the $D$-dimensional isotropic harmonic oscillator up to energy $E$, $\sum_i^D\frac{x_i^2}{2}+\frac{p_i^2}{2}$, ($\omega_i=m_i=1$) is a sphere with radius $R=\sqrt{2E}$ and volume $V=\frac{\pi^DR^{2D}}{D!}=\frac{\pi^D(2E)^{D}}{D!}$. This gives volume $v=\pi R^2$ 1-dimension and volume 
\begin{eqnarray}
{V}=\frac{v^D}{D!}
\label{V}
\end{eqnarray}
in $D$-dimensions (Fig(\ref{volume})). 

In 1-d, the relation between states and volume is $g=\frac{v}{h}$. Thus, we obtain a formula for the scaling of the total number of states similar to that for the phase space volume: 
\begin{eqnarray}
G=\frac{V}{h^D}=\frac{g^D}{D!}.
\label{GV}
\end{eqnarray}
Because of the $D!$ in Eq.(\ref{GV}), $G$ increases slower than exponentially. In fact according to Eq.(\ref{GV}) $G$ {\it decreases} for $D>g$; however as we will show in Section {\ref{sec:number_of_states} this effect is an artifact of using the semiclassical formula outside of its domain of applicability.   

\begin{figure}[h]
\begin{center}
\includegraphics [width=3cm]{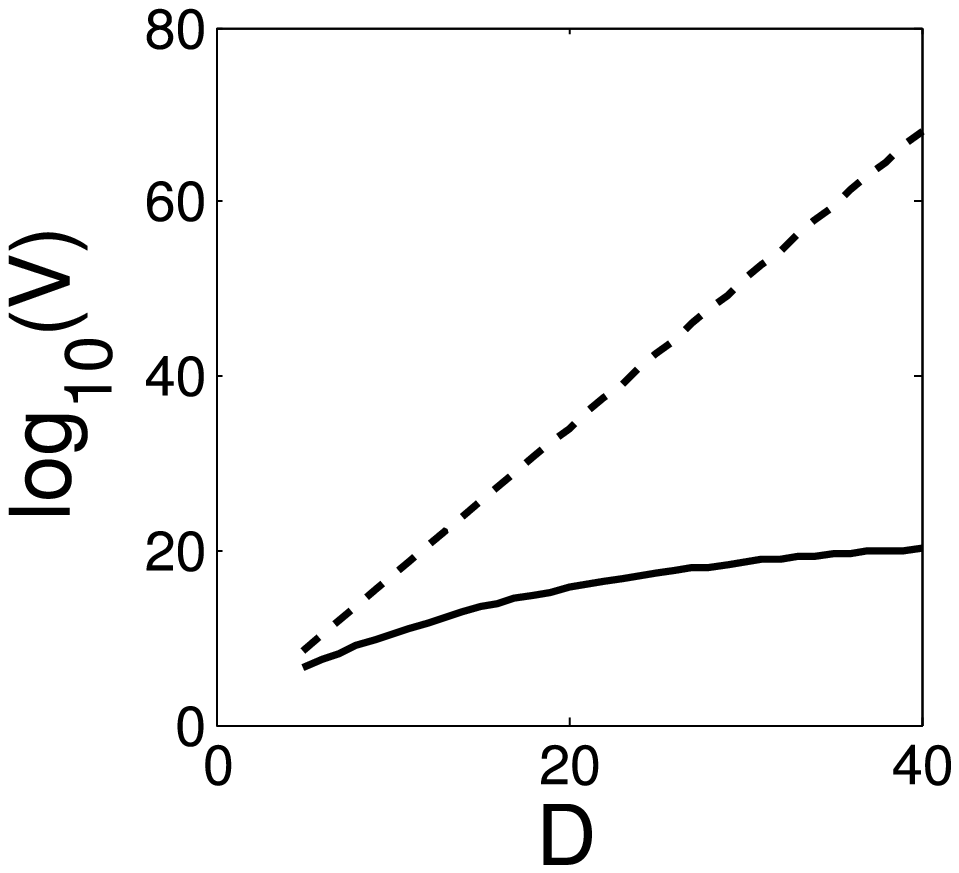}
\end{center}
\begin{center}
\caption{\footnotesize{Log of the volume of the classical phase space below $E=8$ as a function of dimensionality (solid) compared with exponential scaling(dashed).}}
\label{volume}
\end{center}
\vspace{-.5cm}
\end{figure}

\subsection{Ratio between phase space volume and box} 
\label{sec:efficiency}
\subsubsection{Harmonic potentials}
The efficiency of the pvN method relative to the FGH in the classical limit is (the inverse of) the ratio of the classical phase space area and the enclosing box.  For a 1-d harmonic oscillator, this is the ratio between the area of a circle and the square surrounding it, given by: $s=\frac{\pi R^2}{(2R)^2}=\frac{\pi}{4}$.
In $D$-dimensions, this ratio is $S=\frac{s^D}{D!}=\frac{(\pi/4)^D}{D!}$. Because of the $D!$, the relative efficiency (the inverse of $S$) increases faster than exponentially with dimension $D$.

\subsubsection{1-d anharmonic potentials}
For the harmonic oscillator, the packing parameter $s=\frac{\pi}{4}\approx0.7854$  is relatively close to 1. However
most chemically relevant potentials have classical phase space shapes that are packed in a much less compact way (Fig.(\ref{phase_space})). For example, for the Morse oscillator shown in Fig.\ref{phase_space}(b) $s=0.4638$; for the Coulomb potential shown in Fig.\ref{phase_space}(c) $s=0.0321$. 
\begin{figure}[h]
\begin{center}
\includegraphics [width=7cm]{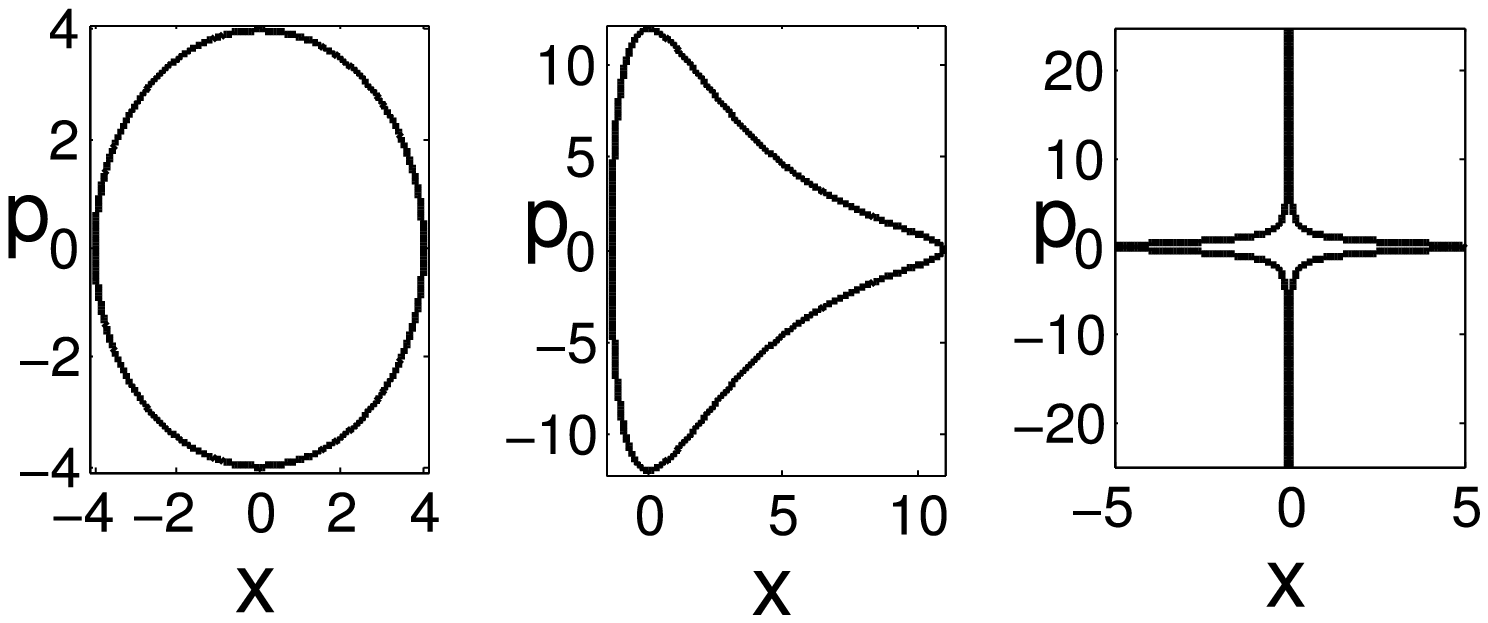}
\end{center}
\begin{center}
\caption{\footnotesize{ Phase space area of (a) harmonic, (b) Morse and (c) Coulomb potentials.}}
\label{phase_space}
\end{center}
\vspace{-0.9cm}
\end{figure}
\subsubsection{$D$-d anharmonic potentials}
For anharmonic potentials, not only is $s<\frac{\pi}{4}$, but the multidimensional classical phase space volume scales more slowly than in Eq.(\ref{V}). This can be understood by visualizing figures such as Figs. \ref{phase_space}(b)-(c) in multidimensions, and is confirmed numerically using Monte Carlo integration (see Fig.(\ref{scale_morse})); in Section \ref{sec:number_of_states} we rationalize this behavior in a different way using state counting arguments. As a consequence, the ratio of phase space volumes $S_{\rm AHO}$ is even smaller than $S_{\rm HO}=\frac{s^D}{D!}$ (and the efficiency of pvN is therefore larger):
\begin{eqnarray}
S_{\rm{HO}}=\frac{s^D_{\rm{HO}}}{D!}>\frac{s^D_{\rm{AHO}}}{D!}>S_{\rm{AHO}}.
\end{eqnarray}
Since the harmonic efficiency (the inverse of $S_{\rm{HO}}$) increases faster than exponentially with dimension $D$, clearly the anharmonic efficiency (the inverse of $S_{\rm{AHO}}$) does as well.  
\begin{figure}[h]
\begin{center}
\includegraphics [width=4cm]{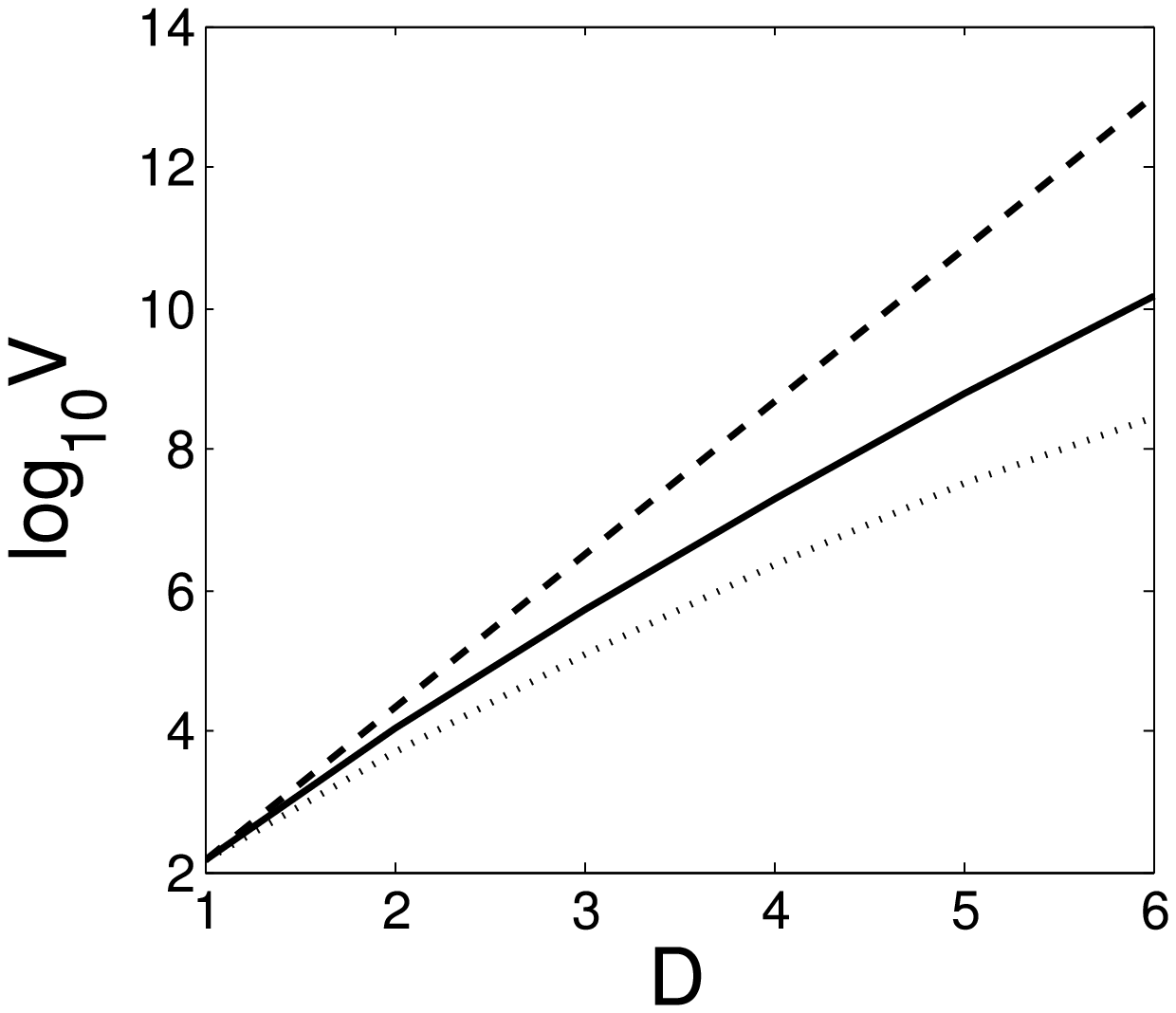}
\end{center}
\begin{center}
\caption{\footnotesize{Log of the phase space volume $V$ vs. $D$ for a $D$-dimensional sum of Morse potentials, calculated using Monte Carlo integration (dots) and using Eq.(\ref{V}) (solid). Exponential scaling is demonstrated by the dashed line.}}
\label{scale_morse}
\end{center}
\vspace{-.8cm}
\end{figure}

\subsection{Number of states for $D$-dimensional potential} \label{sec:number_of_states}
In Section \ref{sec:ho_volume} we showed that for the $D$-dimensional isotropic harmonic oscillator, in the semiclassical limit, the number of states below energy $E$ is given by $G=g^D/D!$ We now reexamine this result starting from a fully quantum treatment. For a 1-d oscillator the number of states below energy $E$ is $g\approx\frac{E}{\hbar\omega}$. For a $D$-dimensional isotropic harmonic oscillator the number of states below $E$ is \textit{not} $g^D$, since some combinations give energy above $E$. Computing $G$ is identical to the number of ways to distribute no more than $g$ indistinguishable balls (quanta of energy) into $D$ distinguishable boxes (oscillators). The answer is given by \cite{Forst}:
\begin{eqnarray}
G=\sum_{i=0}^{g}\frac{i+D-1}{i!(D-1)!}=\frac{(g+D)!}{g!D!}. 
\label{G}
\end{eqnarray}
We now draw attention to two points about Eqn. \ref{G}.\\
1) Consider the case $D=2$ with $g \gg 2$. Then Eqn. \ref{G} becomes 
\begin{eqnarray}
G=\frac{(g+2)(g+1)}{2} \approx \frac{g^2}{2}.
\label{eq:2d_states}
\end{eqnarray}
The factor $g^2$ in the numerator has its origin in the tensor product Hilbert space of the two truncated harmonic oscillators.  The factor of 2 in the denominator comes from the fact that about half of the states in the tensor product Hilbert space have energy $E_1 + E_2 > E$. For a $D$-dimensional system the fraction of tensor product states with $\Sigma_{i=1}^D E_i < E$ is approximately $1/N!$

In the case of an anharmonic potential such as the Morse oscillator, the gaps between energy levels are not equal: the higher energy states are spaced more closely together. As a consequence, for a larger fraction of combinations than in the harmonic case $\Sigma_{i=1}^D E_i > E$ and the scaling of $G$ is slower than in Eq.(\ref{GV}) (Fig.(\ref{scale_morse})). \\
2) If we assume $g \gg D$, Eq.(\ref{G}) becomes 
\begin{eqnarray}
G=\frac{g^D}{D!},
\label{G2'}
\end{eqnarray}
which is the same result as Eq.(\ref{GV}). However, in the opposite limit, $D \gg g$, we obtain 
\begin{eqnarray}
G=\frac{D^g}{g!},
\label{eq:D>g}
\end{eqnarray}
showing that for large enough $D$ the scaling $D$ is actually polynomial.  Equations \ref{G}-\ref{eq:D>g} are plotted in Fig. \ref{fig:number_of_states}.  The true scaling with $D$ is clearly subexponential. Equation \ref{G2'} is seen to be a valid approximation for $g \gg D$, although the turnover at large $D$ is an artifact of using the semiclassical formula outside of its range of validity.
\begin{figure}[h]
\begin{center}
\includegraphics [width=4cm]{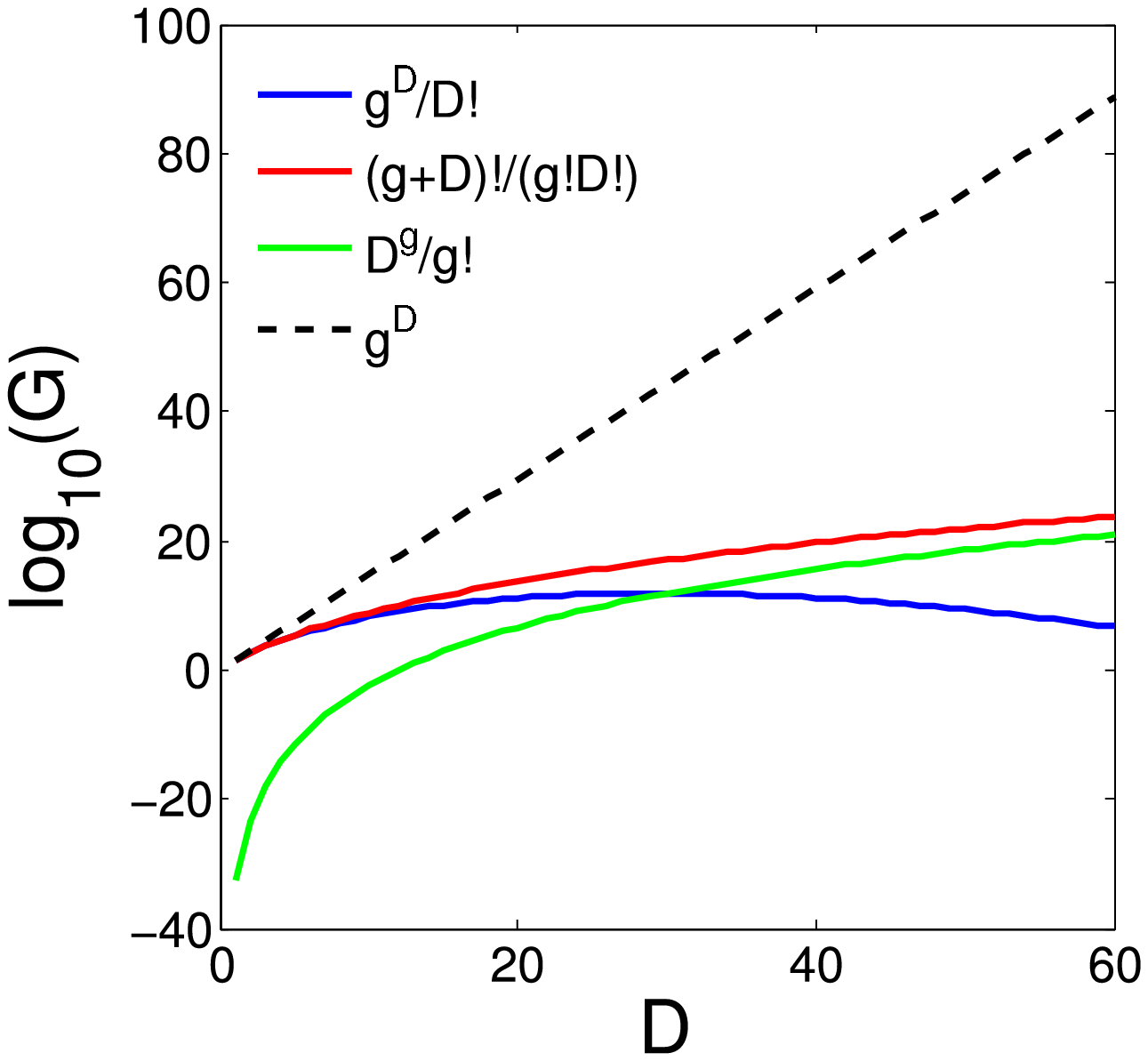}
\end{center}
\begin{center}
\caption{\footnotesize{Log of the number of states $G$ that have enegy below $E=g\hbar\omega$, ($g=30$) vs. $D$ for a $D$-dimensional harmonic potential. The figure shows the exact expression (red), the approximations for $g\gg D$ (blue), for $D\gg g$ (green) and a comparison with exponential scaling (dashed).}}
\label{fig:number_of_states}
\end{center}
\end{figure}